\documentclass[doublecol]{epl2} 
\usepackage{graphicx}
\usepackage{dcolumn,xcolor,ulem}
\usepackage{amssymb,amsmath,amsfonts,amsxtra,amsthm}
\usepackage{bm}
\usepackage[latin1]{inputenc}

\title{Criterion for purely elastic Taylor-Couette instability in the flows of shear-banding fluids\\}

\author{M.A.~Fardin\inst{1,2} \and T.J.~Ober\inst{2} \and C.~Gay\inst{1} \and G.~Gr\'egoire\inst{1}  \and G.H.~McKinley\inst{2} \and S.~Lerouge\inst{1}\thanks{\mbox{\emph{E-mail~address}:~sandra.lerouge@univ-paris-diderot.fr}}}
\shortauthor{M.A.~Fardin \etal}

\institute{                    
  \inst{1} Laboratoire Mati\`ere et Syst\`emes Complexes, CNRS UMR 7057. Universit\'e Paris Diderot, 10 rue Alice Domont et L\'eonie Duquet, 75205 Paris C\'edex 13, France\\
  \inst{2} Department of Mechanical Engineering. Massachusetts Institute of Technology, 77 Massachusetts Avenue, MA 02139-4307 Cambridge, USA
}

\abstract{In the past twenty years, shear-banding flows have been probed by various techniques, such as rheometry, velocimetry and flow birefringence. In micellar solutions, many of the data collected exhibit unexplained spatio-temporal fluctuations. Recently, it has been suggested that those fluctuations originate from a purely elastic instability of the flow. In cylindrical Couette geometry, the instability is reminiscent of the Taylor-like instability observed in viscoelastic polymer solutions. In this letter, we describe how the criterion for purely elastic Taylor-Couette instability should be adapted to shear-banding flows. We derive three categories of shear-banding flows with curved streamlines, depending on their stability.}

\begin{document}

\maketitle

``The stability of viscous liquids contained between two rotating cylinders'' of radii $R_i$ and $R_o$--or Taylor-Couette (TC) flow--is the benchmark problem for instability of flows with curved streamlines. It was the title of a seminal paper by G.I. Taylor in 1923~\cite{Taylor23}, wherein Taylor showed that the purely annular flow eventually becomes unstable. Above a critical rotation speed, a secondary vortex flow sets in, with periodicity $\lambda\sim d$ along the vorticity direction, where $d\equiv R_o - R_i$. The original study by Taylor concerned simple incompressible Newtonian fluids. But many fluids are non-Newtonian and exhibit viscoelastic contributions to the stress~\cite{Larson99}. In 1990, Larson, Shaqfeh and Muller showed that the TC flow of polymer solutions could also become unstable to a Taylor-like instability~\cite{Larson90}. The kinematics of the unsteady flow are roughly similar to those of the Newtonian case, \textit{i.e.} after a critical threshold, Taylor vortices appear, but the destabilizing mechanisms are very different, depending on two different kind of non-linearities. \\
It is well known that Newtonian fluids can exhibit increasingly unstable flows for large values of the Reynolds number. When only the inner cylinder is rotating, the Reynolds number depends on the rotation rate of the inner cylinder $\Omega_i$, such that $Re \equiv \frac{\Omega_i R_i d}{\nu}$, where $\nu$ is the kinematic viscosity of the fluid~\cite{Chandrasekhar81}. In a simple Newtonian fluid, the constitutive relation is the simple linear relation between stress and shear rate. Then, the only non-linearity in the equations of motion comes from the advective term on the velocity $ ({\bf{v.\nabla}})\bf{v}$, in the equation of motion to ensure consistency between Eulerian and Lagrangian descriptions of fluid motion. The Reynolds number $Re$ is linked to the relative magnitude of this term with respect to the dissipation terms~\cite{Chandrasekhar81}. \\
In polymer solutions, and in many non-Newtonian fluids, the primary non-linearity usually comes from the constitutive relation rather than the momentum balance. The constitutive equation is dynamical, $i.e.$ it relates to the stress relaxation dynamics and typically includes a convected derivative on the stress $\bf{T}$~\cite{Larson99}. In this convected derivative, consistency between Eulerian and Lagrangian descriptions requires a convective term, now applied on the stress $({\bf{v.\nabla}})\bf{T}$, and material frame independence requires additional terms of similar dimensionality ${\bf{\nabla v}}.\bf{T}$~\cite{Larson99,Larson90}. The dimensionless group linked to the magnitude of those new non-linear term is the Weissenberg number $Wi \equiv \tau \dot\gamma$, where $\dot\gamma \equiv \frac{\Omega_i R_i}{d}$ is the typical shear rate in the flow and $\tau$ is the stress relaxation time~\cite{Dealy10}. The similarity between $Wi$ and $Re$ is more apparent by defining $Re$ as a function of the viscous diffusion time $\tau_{vd} \equiv \frac{d^2}{\nu}$, $Re = \tau_{vd} \dot\gamma$~\cite{Groisman98}. $Re$ controls the magnitude of the `inertial non-linearity', while $Wi$ controls the magnitude of the `elastic non-linearity'~\cite{Morozov07}. In general, both $Re$ and $Wi$ are non-zero, but in many practical cases for polymer solutions and melts, the elasticity number $\mathcal{E}\equiv \frac{Wi}{Re}=\frac{\tau}{\tau_{vd}}$ is large, leading to negligible inertial effects.\\
In the simplest TC flow, where only the inner cylinder is rotating, and in the small gap limit, \textit{i.e.} $d\ll R_i$, there exist two dimensionless groups, one relevant to the `purely inertial TC instability' $\Sigma_i \equiv \sqrt{\Lambda} Re$--derived and observed by Taylor~\cite{Taylor23}--and one to the `purely elastic TC instability' $\Sigma_e \equiv \sqrt{\Lambda} Wi$--derived and observed by Larson \textit{et al.}~\cite{Larson90}. Here, $\Lambda \equiv \frac{d}{R_i}$ is the geometrical ratio linked to the streamline curvature, necessarily finite for the instability to be linear~\cite{Taylor23,Morozov07}. Note that the Taylor number is usually defined as $Ta\equiv \Sigma_i^2$~\cite{Chandrasekhar81}. In the purely inertial case, the flow becomes unstable for $\Sigma_i >m'$. In the purely elastic case, the flow becomes unstable for $\Sigma_e >m$. Both $m'$ and $m$ are coefficients of order unity, with precise values that depend on the boundary conditions~\cite{Chandrasekhar81,Khayat99}.\\
In this letter, we extend the expression of the instability criterion for viscoelastic `shear-banding flows'. Shear-banding is yet another curious but ubiquitous phenomenon occurring in complex fluids~\cite{Cates06,Lerouge09}. When a fluid material is sheared, the strain rate gradient can be very large in narrow zones of the sample. Adjacent domains of markedly different strain rates are identifiable. This phenomenon has been observed in a variety of systems but in this letter, we focus on the steady shear-banding phenomenon of `wormlike micelles'~\cite{Cates06,Lerouge09}. Entangled wormlike micellar solutions are model rheological fluids due to their linear Maxwellian behaviour for small deformations, characterized by a single relaxation time $\tau$ and elastic modulus $G_0$~\cite{Larson99}. Furthermore, the robustness of their shear-banding behaviour makes them attractive for the general study of banding phenomena in complex fluids~\cite{Rehage91}. Roughly speaking, a shear-banding flow is reminiscent of a first order phase transition. Above a lower critical Weissenberg number $Wi_l$, the shear stress plateaus. Then, until a second higher critical Weissenberg number $Wi_h$, the flow is inhomogeneous, split in two bands with local Weissenberg numbers $Wi_l$ and $Wi_h$. To leading order, for $Wi \in [Wi_l, Wi_h]$, an increase in the value of the macroscopic Weissenberg $Wi$ only increases the proportion $\alpha\in[0,1]$ of the high $Wi$ band, following a `lever rule' $Wi\simeq\alpha Wi_h +(1-\alpha)Wi_1$~\cite{Cates06,Lerouge09}. This scenario has been roughly confirmed experimentally with various techniques  \textit{e.g.} pure viscometry, velocimetry, birefringence, \textit{etc.}~\cite{Lerouge09} but many fluctuating behaviours were observed in all the gathered data~\cite{Lerouge09}. \\
In a series of recent experiments~\cite{Lerouge06,Lerouge08,Fardin09,Fardin10} we have shown that the interface between the bands undulates due to an underlying secondary vortex flow that is mainly localised in the high Weissenberg number ($Wi_h$) band. Elastic instabilities similar to the one observed in polymer solutions could be the source of many of the observed spatio-temporal fluctuations. This rationale was reinforced by a recent linear stability analysis of the diffusive Johnson-Segalman (dJS) model~\cite{Fielding10}, a viscoelastic constitutive model widely used to study shear-banding flows~\cite{Fielding07}. In this letter we wish to rationalize experimental data on the shear-banding TC flow of wormlike micelles by determining the appropriate form of the instability criterion in the case of shear-banding flow.

\section{Rheological and geometric scaling of purely elastic flow instabilities}
In the introduction, we discussed the TC instability of the purely Newtonian fluid and the purely elastic fluid, which are both idealizations that facilitate our analysis, but which capture only the behaviour of very specific fluids. In general, however, non-Newtonian fluids can exhibit other attributes such as a Newtonian solvent contribution to the stress, a spectrum of relaxation times instead of a single relaxation time $\tau$, and/or `shear thinning', \textit{i.e.} a decreasing viscosity with increasing shear rate~\cite{Larson99}. Experiments conducted on such non-Newtonian fluids have documented the effects of such fluid rheology on the elastic TC instability~\cite{Larson94}. To rationalize these observations as well as to generalize the elastic instability criterion to different kinds of flows with curved streamlines, McKinley and coworkers established a general criterion for elastic instabilities~\cite{Mckinley96,Mckinley96b}. If $Re\simeq 0$, then, viscoelastic fluids are unstable if $\frac{N_1}{T_{xy}}\frac{\ell}{\mathcal{R}}> m^2$, where $N_1\equiv T_{xx} -T_{yy}$ is the first normal stress difference~\cite{Larson99}, $T_{xy}$ is the shear stress, $\ell$ is the characteristic distance over which perturbations relax along a streamline~\cite{Mckinley96}, and $\mathcal{R}$ is the characteristic radius of curvature of the streamlines. For a purely viscoelastic fluid, $\ell \equiv U \tau \sim \Omega_i R_i \tau = Wi~d$, $\mathcal{R} \sim R_i$ and $\frac{N_1}{T_{xy}}=\frac{N_1}{T_{\theta r}} \sim Wi$~\cite{Mckinley96} and we recover the criterion of  Larson \textit{et al.} for the purely elastic instability: $\Lambda Wi^2 > m^2 \Leftrightarrow \sqrt{\Lambda} Wi > m$ ~\cite{Larson90}. \\
In turn, framed with respect to the general criterion derived by McKinley \textit{et al.}~\cite{Mckinley96,Mckinley96b}, our goal is to determine the functional form of the dimensionless ratio $\frac{N_1}{T_{xy}}\frac{\ell}{\mathcal{R}}$ in terms of measurable quantities in the case of shear-banding flows. By analogy with polymer solutions, we would expect that this ratio can be expressed in terms of a relevant geometric ratio and an appropriately-defined Weissenberg number.

\section{Effective gap}
\begin{figure}
\centering
\includegraphics[trim = 0mm 0mm 0mm 0mm,width=6.5cm,clip]{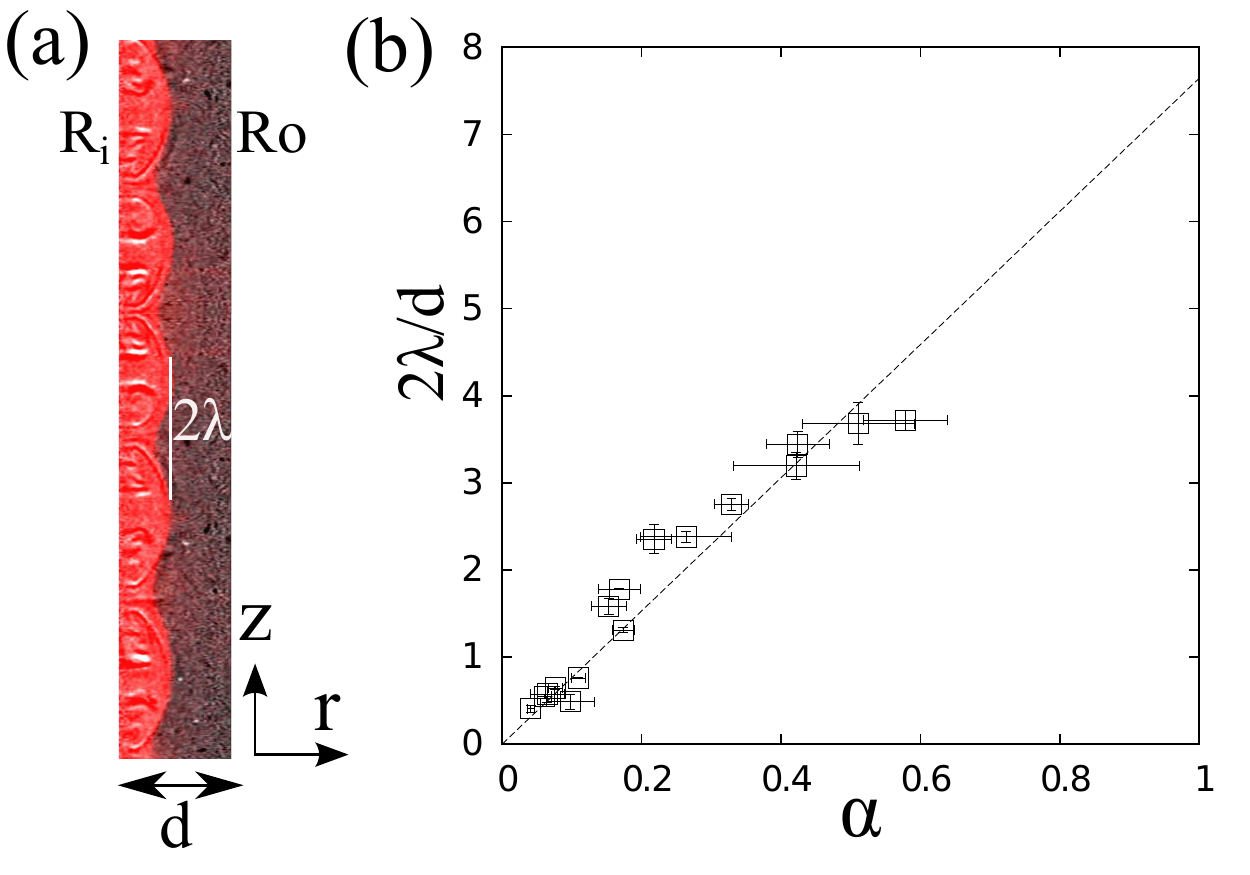}
\caption{Effective gap scaling. (a) Overlay of two visualization techniques showing the secondary vortex flow in the high $Wi$ band for $\alpha\simeq 0.4$~\cite{Fardin09}. (b) Wavelength scaling, following $\lambda=n \alpha d$, with $n=3.8\pm 0.1$. For $\alpha>0.6$, the spatio-temporal dynamics of the vortex flow do not allow us to extract a single wavelength~\cite{Lerouge08,Fardin09}. For $\alpha<0.05$, the size of the band is smaller than our spatial resolution.      	
\label{Fig1}}
\end{figure}
The relevant geometric ratio can indeed be inferred from experiments through the notion of an \textit{effective gap}. In our recent experiments~\cite{Fardin09,Fardin10}, we recognized that the vortices were mainly localized in the high $Wi$ band, and that each interfacial wavelength between the bands corresponded to a pair of counter-rotating vortices~\cite{Fardin09}, as illustrated in Fig. \ref{Fig1}a. In our previous publications, we had noticed that the wavelength increases upon increase of the global shear rate, so one could infer the scaling $\lambda/d \sim Wi$~\cite{Lerouge08,Fardin09}. Then, by combining this scaling and the lever rule we can establish that $\lambda = n \alpha d$ instead of $\lambda=n d$, where $n$ is a number of order unity, whose precise value depends on the boundary conditions. The extent of the high $Wi$ band acts as the effective gap. Increasing the global $Wi$ increases $\alpha$ and so increases $\lambda$. The validity of this scaling is shown on Fig.~\ref{Fig1}b by re-plotting $2\lambda/d$, \textit{i.e.} twice the wavelength of vortices, against $\alpha$ instead of $Wi$~\cite{Lerouge08}.

\section{Local Weissenberg number}
As explained in the introduction, in a shear-banding flow, the global value of $Wi$ is not a good measure of the local Weissenberg number in the parts of the flow that are unstable. Instead, the dimensionless group relevant to the flow instability is the local value of $Wi_h$ in the high shear rate band. In the instability criterion, one must replace $Wi$ by $Wi_h$. Accordingly, the criterion for elastic instabilities in shear-banding flows should involve the term
\begin{equation}
\Sigma^* = \sqrt{\alpha \Lambda} Wi_h 
\label{effgap}
\end{equation} 
It has been observed in experiments that increasing the concentration ($c$) of surfactant, or decreasing the temperature ($\theta$) tends to increase the value of $Wi_h$. This fact is illustrated in Fig.~\ref{Fig2}a in the flow curves of two different surfactant systems~\cite{Berret97,Cappelaere97}. Note that as $c$ increases, the dimensionless stress plateau decreases and its range of Weissenberg numbers increases. In particular, $Wi_h$ shifts to higher values. For the most concentrated solutions, viscometric measurements had to be aborted because the sample was ejected from the rheometer. We believe that this phenomenon is due to an instability of the free surface of the system, driven by the underlying bulk viscoelastic instability. However, we also note that the instability of the free surface could be triggered by second normal stress differences~\cite{Skorski11}. From Eq.~(\ref{effgap}), we note that solutions of high $c$ or at low $\theta$ are more likely to be unstable, owing to the larger values of $Wi_h$.\\

\section{The case of dJS}
\begin{figure}
\centering
\includegraphics[trim = 7mm 5mm 0mm 0mm,width=6cm,clip]{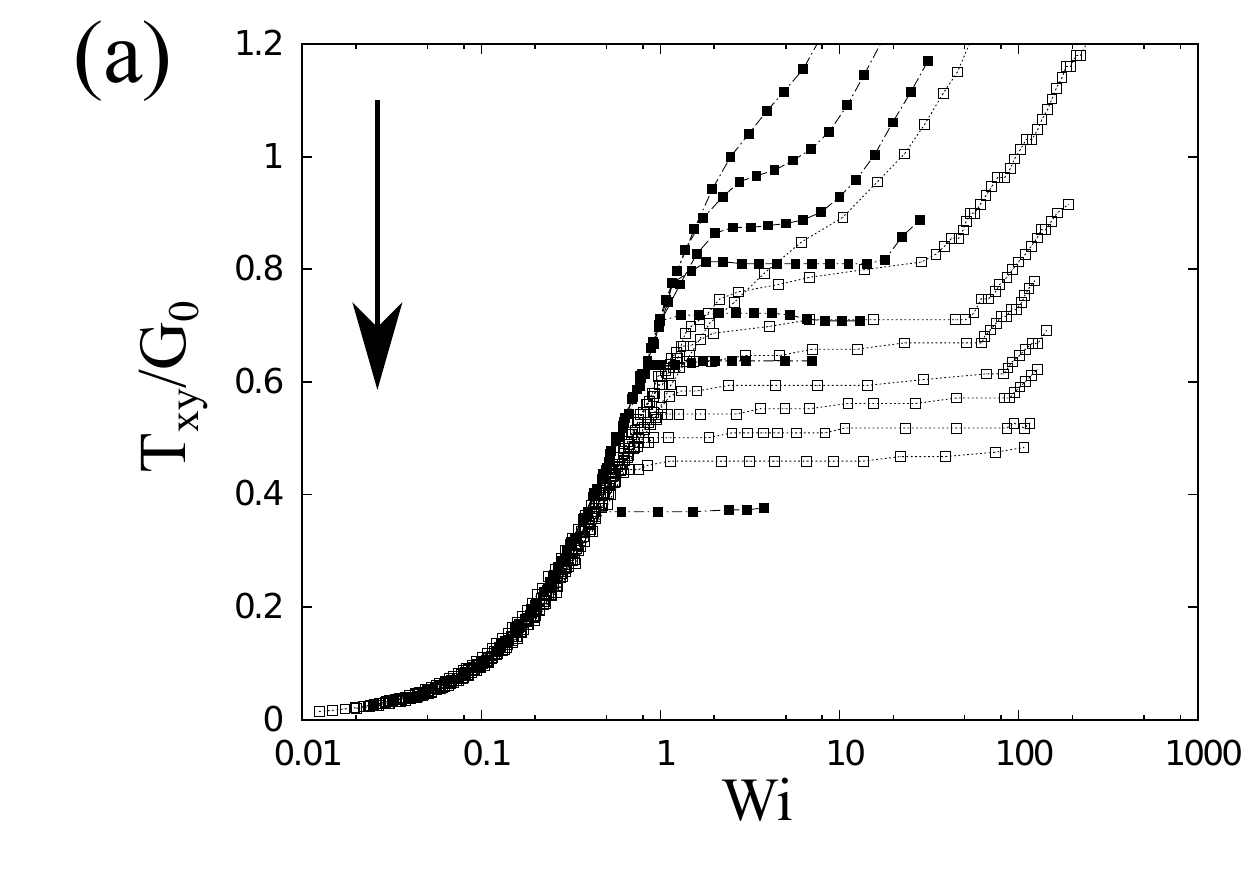}
\includegraphics[trim = 2mm 15mm 5mm 0mm,width=6.5cm,clip]{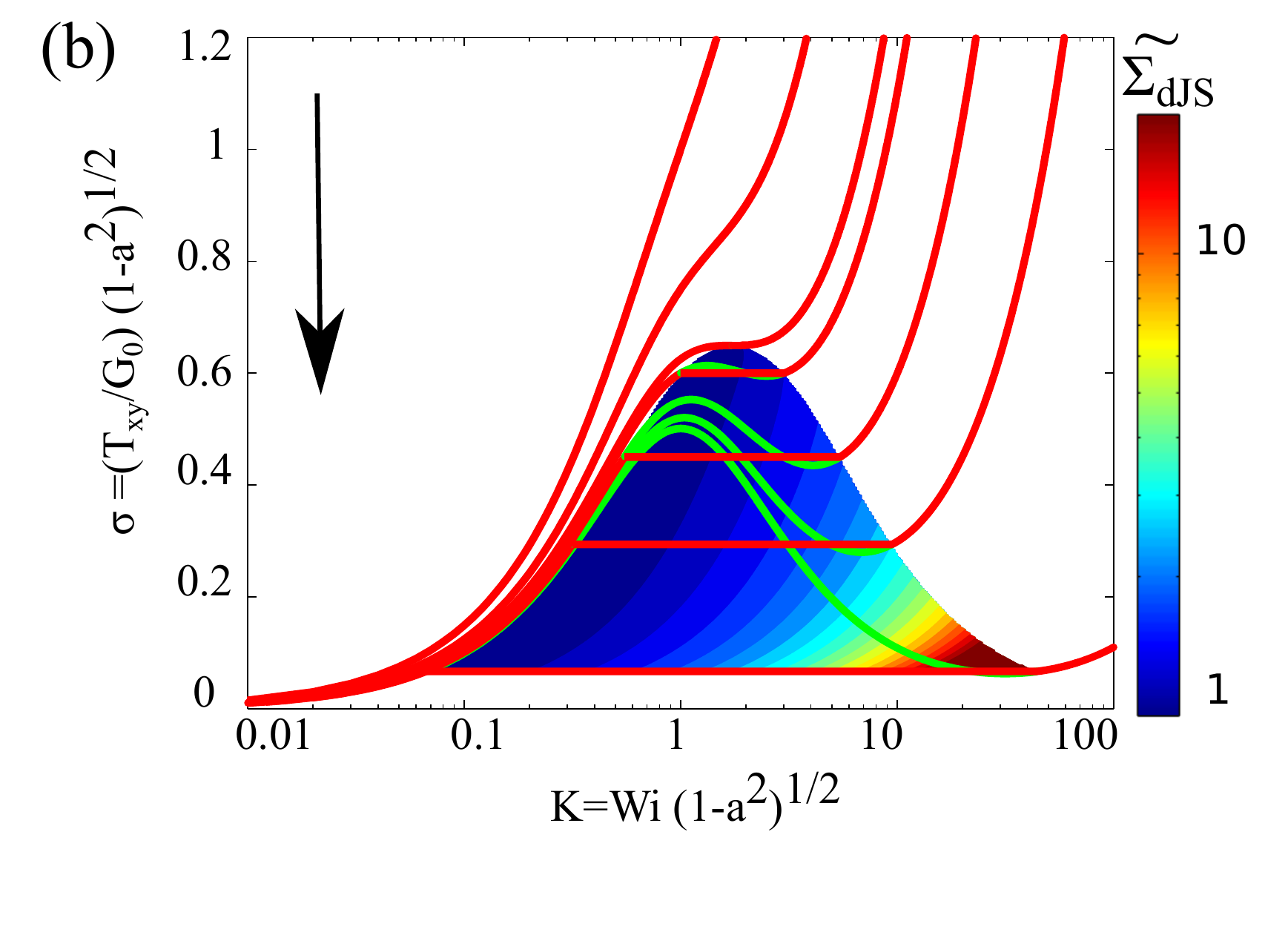}
\caption{Experimental and theoretical ``flow-phase diagrams''~\cite{Berret97}. (a) Open symbols: Measured dimensionless flow curves for varying [CTAB]=3,7,10,12,15,17,18,22wt.$\%$ at fixed [NaN0$_3$]=0.3M (replotted data adapted from Cappelaere \textit{et al.}~\cite{Cappelaere97}, permission from Springer). Closed symbols: Measured dimensionless flow curves for varying [CPCl+0.2NaSal]=2,4,6,8,10,12,21wt.$\%$ (courtesy of Berret \textit{et al.}~\cite{Berret97}). The arrow points in the direction of higher $c$ or lower $\theta$~\cite{Berret97}. In both cases, measurements were done using a cone-and-plate device. The flow curves of the two systems do not overlap, even when the stress and the shear-rate are scaled with $G_0$ and $\tau$ respectively, \textit{i.e.} in the framework of the dJS model, the two systems have a different value of the coefficient `\textit{a}'. (b) Analytical dimensionless flow curves obtained for the dJS model in simple shear~\cite{Sato10}. The different flow curves are obtained for varying	$\eta$. The color map gives the value of the scaled dimensionless criterion $\widetilde{\Sigma_{dJS}}\equiv \Sigma_{dJS} \sqrt{\frac{1-a^2}{\Lambda}}$. The arrow points in the direction of lower $\eta$. 	
\label{Fig2}}
\end{figure}
So far, we have suggested a new relevant dimensionless group for elastic instabilities in shear-banding flows, without appealing to any particular rheological model. To reinforce our argument, we can investigate the form of the instability criterion for the diffusive Johnson-Segalman (dJS) model, which has been widely used to study shear-banding flows~\cite{Fielding07}. Recently, it has even been used in numerical simulations confirming the presence of a secondary vortex flow triggered by a bulk viscoelastic instability in the high $Wi$ band~\cite{Fielding10}. In this model, the stress is taken as the sum of a `polymeric' part $\bf{T^p}$ and a `solvent' part $\bf{T^s}$. The total viscosity of the fluid is the sum of a polymeric and solvent part $\eta_0=\eta_p + \eta_s$, with the zero shear rate value of the polymeric viscosity given by $\eta^0_p\equiv \tau G_0$. The polymeric stress varies non-monotonically with imposed shear rate and goes to zero at large $Wi$, such that $\eta_s$ is the asymptotic value of viscosity for $Wi\to\infty$. \\
To evaluate the expression $\frac{N_1}{T_{xy}}\frac{\ell}{\mathcal{R}}$, we need an analytic expression for the stress ratio in the high $Wi$ band. Let us symbolize this ratio by $[\frac{N_1}{T_{xy}}]_h$. In the small gap limit, we can assume that the stress profile across the gap is close to the profile in a plane Couette geometry. We can then use the inhomogeneous plane Couette solution recently derived by Sato \textit{et al.}~\cite{Sato10}.\\
In plane Couette flow of the dJS model, it is common to express the total shear stress as $T_{xy}= \frac{G_0}{\sqrt{1-a^2}} \sigma $, and the first normal stress difference as $N_1= \frac{2G_0}{1-a^2} N$~\cite{Sato10}. The parameter $a$ is the `slip parameter' of the dJS equation. Shear-banding happens if $|a|\neq 1$ and $\eta\equiv \frac{\eta_s}{\eta^0_p}<\frac{1}{8}$~\cite{Fielding07,Sato10}. $N$ is a dimensionless normal stress difference and $\sigma$ is a dimensionless total shear stress. In plane Couette flow, the momentum balance imposes that $\sigma$ is a constant across the gap, but $N(y)$ is a function of the position $y$ in the gap~\cite{Sato10}. For steady flow in the shear-banding regime, Sato \textit{et al.} established that
\begin{align}
\sigma &= 3\frac{\sqrt{\eta -\eta^2}}{\sqrt{2}}\\
N & =KS=K(\sigma-\eta K)
\label{Neq}
\end{align} 
where $K(y)\equiv \sqrt{1-a^2} Wi(y)$ and $S(y)\equiv \sqrt{1-a^2} \frac{T^p_{xy}(y)}{G_0}$ are respectively a dimensionless shear rate and a polymeric shear stress, both functions of the position in the gap. In dimensionless form, the addition of the polymeric and solvent shear stress is expressed by $\sigma\equiv S(y)+\eta K(y)$~\cite{Sato10}. In the shear-banding regime, Sato \textit{et al.} have found an analytic solution for the dimensionless shear rate profile $K(y)$ that follows a hyperbolic tangent profile between $K_l$ and $K_h$~\cite{Sato10}, with $K_l<K_h$ given by  
\begin{align}
& K_{l} = \frac{\sqrt{1/\eta - 2} - \sqrt{1/\eta -8}}{\sqrt{2}} \\
& K_{h} = \frac{\sqrt{1/\eta - 2} + \sqrt{1/\eta -8}}{\sqrt{2}} 
\label{Kheq}
\end{align} 
In the high shear rate band, $K\simeq K_h=\sqrt{1-a^2} Wi_h$. Thus from eqs.~(2), (\ref{Neq}) and (\ref{Kheq}) we can obtain the following expressions
\begin{align}
\Big[\frac{N_1}{T_{xy}}\Big]_h &= \Big[\frac{N}{\sigma}\Big]_h ~\frac{2}{\sqrt{1-a^2}} \\
							&= \frac{K_h(\sigma-\eta K_h)}{\sigma} \frac{2}{\sqrt{1-a^2}}\\
							&= \frac{2}{3} Wi_h \Big(2 -   \sqrt{\frac{1-8\eta}{1-2\eta}}\Big)												
\label{stressratioWih}
\end{align}  
Then, overall, if we set $\ell\sim Wi_h \alpha d$ and $\mathcal{R}\sim R_i$, we get  
\begin{equation}
\Sigma_{dJS} = \sqrt{\alpha \Lambda} Wi_h f(\eta) =\Sigma^* f(\eta) 
\label{dJSscaling}
\end{equation}
Therefore, the result we obtain using dJS is slightly more complex than the naive criterion $\Sigma^*$ since it also depends on the viscosity ratio. For shear-banding we require $\eta<1/8$, so we have $0.7 \lesssim f(\eta) \lesssim 1.3$. This result is indeed not surprising, since we had obtained $\Sigma^*$ in analogy with the purely elastic case derived using the Upper Convected Maxwell model, where $\eta=0$~\cite{Larson90} . In the homogeneous and non shear-banding elastic case, adding a Newtonian solvent also modifies the dimensionless group by the addition of a function $f^\star(\eta)\simeq \sqrt{\frac{2}{1+\eta}}$~\cite{Mckinley96}. \\
Note that the expression for $\Sigma_{dJS}$ can also be expressed in terms of the two dimensionless variables $K$ and $\sigma$. Indeed, from the lever rule, $\alpha=\frac{K-K_l}{K_h-K_l}$, and from eq. (4) and (5), $K_l$ and $K_h$ can be expressed in terms of $\eta$, which can be subsequently expressed in terms of $\sigma$ using eq. (2). Ultimately, one can reach the following equivalent alternative expression for $\Sigma_{dJS}$:
\begin{equation}
\Sigma_{dJS} = \sqrt{\frac{\Lambda}{1-a^2}} \widetilde{\Sigma_{dJS}}(K, \sigma)
\label{dJSscaling}
\end{equation}
where $\widetilde{\Sigma_{dJS}}(K, \sigma)= (2\sqrt\frac{K}{3\sigma}-\sqrt\frac{\sigma}{3K})+\mathcal{O}[\sigma^{3/2}]$ is a function of $K$ and $\sigma$ only, whose precise functional form is a little too cumbersome to be written explicitly. Figure~\ref{Fig2}b. plots the flow curves computed from eqs. (2), (4) and (5)~\cite{Sato10}, together with the magnitude of $\widetilde{\Sigma_{dJS}}$. We can see that as the shear rate is increased, the proportion of the high $Wi$ band increases, the magnitude of the scaled criterion $\widetilde{\Sigma_{dJS}}$ increases and the flow is increasingly prone to instability. By comparing the experimental flow curves in  Fig.~\ref{Fig2}a and the flow curve derived in the case of dJS in Fig.~\ref{Fig2}b one can see that the effect of decreasing the Newtonian solvent contribution $\eta$ to the total stress is similar to the effect of increasing the concentration of surfactant, or decreasing the temperature.

\section{Boundary conditions and classes of unstable shear-banding flows}
\begin{figure}
\centering
\includegraphics[width=7.5cm,clip]{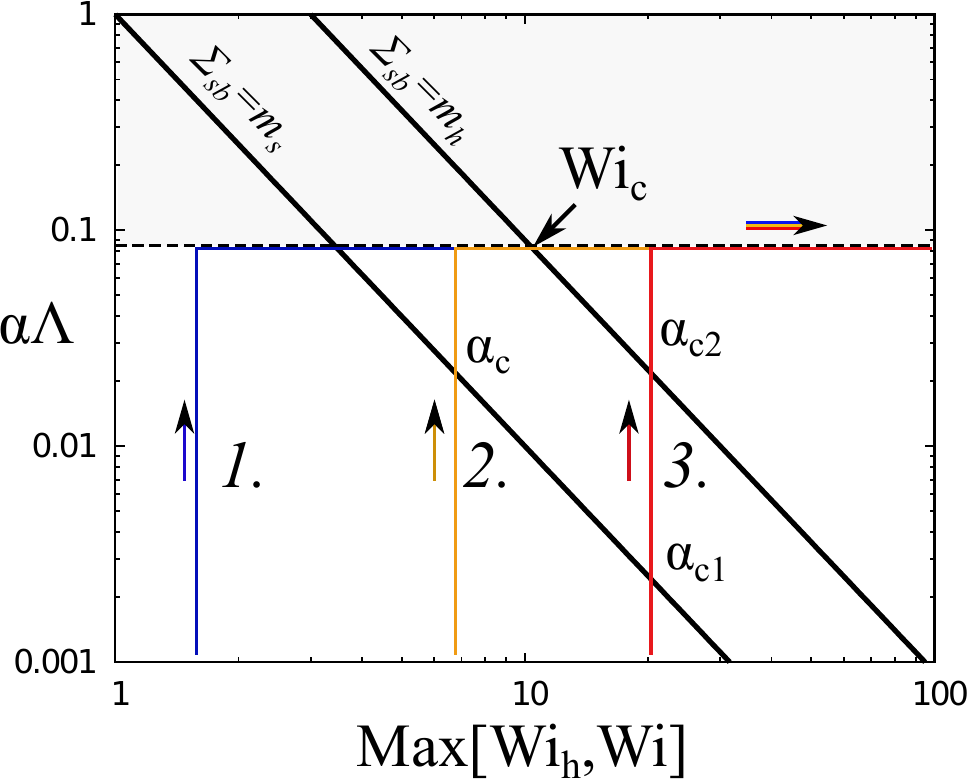}
\caption{Schematic instability diagram in the plane (Max[$Wi_h$,$Wi$],$\alpha\Lambda$). The black lines represent the stability limits for soft and hard boundaries, $\Sigma_{sb}=m_s \Leftrightarrow \alpha\Lambda = (m_s/Wi_h)^2$ and $\Sigma_{sb}=m_h \Leftrightarrow \alpha\Lambda = (m_h/Wi_h)^2$, where we have arbitrarily chosen $m_s=1$ and $m_h=3$.  The dashed black line represent the value of $1\times \Lambda=1.13/13.33$, the maximum curvature corresponding to our recent experiments~\cite{Lerouge06,Lerouge08,Fardin09,Fardin10}. Above this line, the shaded region is inaccessible. The three paths \textit{1.}, \textit{2.} and \textit{3.} illustrate the three possible types of shear-banding. The direction of the arrows represent the path followed by the state of the flow as the global Weissenberg $Wi$ is increased. $\alpha_c$, $\alpha_{c1}$ and $\alpha_{c2}$ are the critical proportions of the high $Wi$ band at which the flow state crosses a stability limit. $Wi_c$ is the threshold at which the type \textit{1.} trajectory becomes unstable for the first time, and at which the type \textit{2.} trajectory becomes unstable after a short relaminarization. 
\label{Fig3}}
\end{figure}
Generally, we expect the relevant dimensionless group for elastic instability in shear-banding flows to be $\Sigma_{sb} \equiv \Sigma^* f^*(\eta)$, where $f^*$ is a function of the ratio between the zero shear and infinite shear viscosities. We expect the specific form of $f^*$ to depend on the constitutive model used to study shear-banding~\cite{Cates06}. Elastic instabilities will generate a secondary vortex flow with wavelength $\lambda=n \alpha d$ for $\Sigma_{sb}>m$. As mentioned already, the precise values of $n$ and $m$ depend on the boundary conditions. Of prime importance are the values of $m$ obtained for `soft' ($m_s$) or for `hard' ($m_h$) boundary conditions~\cite{Fardin10}. Essentially, the `hard' case usually corresponds to a no-slip Dirichlet boundary condition, while the `soft' case usually corresponds to imposing only continuity of the stress, \textit{i.e.} a Neumann boundary condition. In both the purely inertial case~\cite{Chandrasekhar81} and the purely elastic case~\cite{Khayat99}, it is known that $m_s<m_h$. For a banded flow with $Wi\in[Wi_l,Wi_h]$, the interface with the low $Wi$ band acts as a soft boundary for the high $Wi$ band. But for $Wi \geqslant Wi_h$, $\alpha=1$, the flow becomes homogeneous again and the boundary switches from soft to hard. \\
Therefore, for a given geometry, \textit{i.e.} a given value of $\Lambda$, we can use basic Boolean logic to classify shear-banding flows into three possible categories depending only on the value of $Wi_h$: \\
\textit{1.} For sufficiently low $Wi_h$--\textit{i.e.} high $\theta$ and low $c$--the shear-banding flow is stable for any $\alpha$, since $\Sigma_{sb}< m_s$ even for $\alpha=1$. The flow can then become unstable for Weissenberg numbers above a critical value $Wi_c>Wi_h$ as in the case of a regular viscoelastic fluid, \textit{i.e.} following the scaling $\Sigma_i =\sqrt{\Lambda}Wi$.\\
 \textit{2.} For intermediate values of $Wi_h$--\textit{i.e.} intermediate $\theta$ and $c$--the shear-banding flow is unstable above a critical value $\alpha_c$ when $\Sigma_{sb} > m_s$ for $\alpha>\alpha_c$. Then as the imposed shear rate is increased and $\alpha\to1$ the boundary conditions change and the flow is stabilized, because the flow is below the threshold $m_h$. Eventually for $Wi>Wi_c>Wi_h$ the flow becomes unstable again. This case was the one we observed in our recent experiments~\cite{Fardin10}. \\
\textit{3.} Finally, if $Wi_h$ is high enough--\textit{i.e.} for low $\theta$ and high $c$--we have two critical band widths $\alpha_{c1}$ and $\alpha_{c2}$. For $\alpha>\alpha_{c1}$, $\Sigma_{sb} > m_s$. And for $\alpha>\alpha_{c2}$, $\Sigma_{sb} > m_h$. In this case, there is no stabilization for $Wi>Wi_h$. The flow remains unstable, although the spatiotemporal characteristics may change \\
The three possible shear-banding scenarios can be illustrated on a stability diagram in the plane (Max[$Wi_h$,$Wi$],$\alpha\Lambda$), as presented in Fig.~\ref{Fig3}. When the global Weissenberg number $Wi$ is increased above $Wi_l$, the flow state is given by a constant abscissa depending on the value of $Wi_h$ (which is a function of the concentration and temperature of the solution). As $Wi$ increases, the thickness of the high shear rate band $\alpha$ increases and so the state of the flow moves vertically to larger ordinates. Once the entire gap is filled, $\alpha\Lambda$ reaches its maximum depending on the geometry of the chosen TC system. Then, since $Wi>Wi_h$, the state of the flow is given by a constant ordinate $\Lambda$ and moves horizontally as $Wi$ increases. Any flow state with $\alpha\Lambda<\Lambda$ will be stable if below the stability limit $\Sigma_{sb}=m_s$, and unstable if above $\Sigma_{sb}=m_s$ and \textit{a fortiori} if above $\Sigma_{sb}=m_h$. Any flow state with $\alpha\Lambda=\Lambda$ will be stable if on the left of the stability limit $\Sigma_{sb}=m_h$, and unstable otherwise.

\section{Interaction with interface modes}
So far, we have only considered elastic instabilities arising in the bulk of the high $Wi$ band. But there exist other elastic instability mechanisms~\cite{Larson92}. In particular, Fielding has shown that the jump in normal stresses between the bands could generate interfacial modes, even in plane Couette flow~\cite{Fielding07}. In her recent study in TC flow, Fielding suggested that the interfacial and bulk elastic modes lie in two separate regions of the space ($\Lambda$,$N_1|_h$), \textit{i.e.} of the space ($\Lambda$,$Wi_h$)~\cite{Fielding10}. The bulk mode prevails at high $Wi_h$ and high curvature $\Lambda$. The interfacial mode prevails at low $Wi_h$ and low $\Lambda$. Fielding's study would suggest the existence of another unstable region in the lower left corner of the stability diagram sketched in Fig.~\ref{Fig3}. Nonetheless, only axisymmetric perturbations were considered in Fielding's study~\cite{Fielding10}, and the stability analysis was performed for a single value of $\alpha$ and $\eta$. Interfacial and bulk modes may actually interact through non-axisymmetric mechanisms~\cite{Morozov11}.

\section{Wall slip and non-local effects}
We believe that the instability criterion we have derived for shear-banding flows can be a powerful guide to interpret experiments on wormlike micelles. Nonetheless, the criterion is fallible. In particular, we think that two additional phenomena can strongly compromise the validity of our scaling, since both have been shown to be relevant in some experimental situations. In both phenomena the local Weissenberg value in the high shear-rate band may not be equal to the upper boundary of the shear-banding regime on the flow curve. The first phenomenon is wall slip, which has been reported recently and may actually be a common feature of many shear-banding flows~\cite{Lettinga09}. The second phenomenon is geometric confinement. The present scaling may be inadequate if `non-local effects' become dominant~\cite{Masselon08}. `Non-local effects' are apparent in confined geometries when the size $d$ becomes comparable to the typical interfacial width $\xi\sim \mu m$, linked to the stress diffusion coefficient~\cite{Fielding07,Sato10}. Even in a macroscopic geometry with $d\gg \xi$, non-local effects can be important when the lateral extent of one of the bands is very small, \textit{i.e.} $\alpha\simeq 0$ or $\alpha\simeq 1$. Those effects were ignored in the analytic solution for dJS proposed by Sato \textit{et al.} but can actually be derived directly from the dJS equations~\cite{Fardin11}.\\

\indent In summary, we have derived a useful dimensionless criterion to rationalize the onset of secondary flows in the base shear-banding flow of wormlike micelles. The validity of the criterion for the case of dJS could be checked by numerical simulations for various value of the solvent ratio $\eta$, and for a range of gap spacings ($\Lambda$) and Weissenberg numbers. On the experimental side, we are currently undertaking a large study of the stability of shear-banding flows for many different surfactant types, concentrations and temperatures. Preliminary results confirm the existence of the three distinct scenarios that we derived here. Ultimately, the criterion could be extended to other flows with curved streamlines, if the localization and number of bands is known.

%\section{Section title}
%Insert here the text.
%See fig.~\ref{fig.1}, table~\ref{tab.1} and eq.~(\ref{eq.1}).
%See also~\cite{b.a,b.b}.
%\begin{equation}
%\label{eq.1}
%0\neq1
%\end{equation}

%\begin{figure}
%\onefigure{epl-template.eps}
%\caption{Figure caption.}
%\label{fig.1}
%\end{figure}

\acknowledgments
The authors thanks S. Asnacios, O. Cardoso, S.M. Fielding, A.N. Morozov, S.J. Muller and C. Wagner for fruitful discussions. M.A.F. thanks the Fulbright Commission for its support. T.J.O. acknowledges the NSF-GRF for funding.

\end{document}